\documentstyle[twoside,epsfig,fleqn,espcrc2]{article}
\newcommand{\AmS}{{\protect\the\textfont2
  A\kern-.1667em\lower.5ex\hbox{M}\kern-.125emS}}

\title{Infinite coupling magnon theory of quantum Heisenberg magnetic\\ 
       models of spin $s$}

\author{Bang-Gui Liu\address{Institute of Physics, Chinese Academy of 
        Sciences, P O Box 603, Beijing 100080, P R China}
        and 
        Gerd Czycholl\address{Institute for Theoretical Physics, 
         University of Bremen, D--28334 Bremen, Germany}}
       
\begin{document}

\begin{abstract}
An infinity magnon coupling term is introduced into the Holstein-Primakoff 
transformed forms of the Heisenberg ferromagnetic and antiferromagnetic
models of any spin $s$ to rigorously remove the unphysical magnon states. 
This term makes the series expansion of the square root of the magnon 
operators become a finite series of magnon operator products.
Under a simple Hubbard-like approximation our infinite coupling theory
yields much better result than the existing spin wave theories, especially
near transition temperatures.
\end{abstract}

\maketitle

\noindent {\large \it Introduction: }
Quantum Heisenberg ferromagnetic (FM) and antiferromagnetic (AFM) 
models are 
well-accepted models for insulating ferromagnets and antiferromagnets. 
For general parameters one has to
turn to some approximation methods or to numerical work. As for analytical
methods, someone worked directly with the spin operators and their spin 
algebra. 
Anyway, a treatment in terms of the bosonic magnon operators should be
advantageous because of the simpler commutation rules and the Bose
statistics. For this purpose one has to map the spin Heisenberg models 
on spin wave (or magnon) models using the well known 
Holstein-Primakoff\cite{HP40} (HP) or the Dyson\cite{Dyson56} transformation. 
But, on one hand, the Dyson's transformationone breaks the 
conjugate relation of spin operators; and in the Holstein-Primakoff 
transformation one has to treet the square root terms of magnon 
occupation operators. On the other hand, 
the magnon Hilbert space is much larger than the physical Hilbert space. 
For spin $s$, the spin Hilbert space on a single site consists of $2s+1$ 
states but the magnon Hilbert space on a single site is infinite dimensional. 
In 3D ferromagnets there should be only few magnon
excitations at low temperature. But in antiferromagnets, there should be a
substantial number of magnons even at zero temperature because the sublattice
magnetization is less than $s$. The effect of the extra unphysical magnon states
on the physical quantities becomes more serious with increasing temperature.

We introduced an infinite magnon coupling terms into the Holstein-Primakoff
transformed hamiltonians\cite{LiuCzyc} and thereby, for the first time,
completely removed the effect of the extra unphysical states for any spin 
$s$. At the same time this infinite coupling term made the resultant magnon
hamiltonians automatically truncated into finite series of the magnon
operator products without any approximation. Under a simple nonperturbation
approximation our hamiltonians yielded much better results than the existing
spin wave theories.\\

\noindent {\large\it Infinite coupling magnon term: }
We began with the standard isotropic Heisenberg FM and AFM
hamiltonians of any spin $s$. Since Dyson transformation
breaks the conjugate relation of spin operators,
we chose HP transformation to transform the spin
operators into the magnon operators. 
\begin{equation}\begin{array}{l}
S_i^{-}=a_i^{\dagger }\sqrt{2s-n_i},~~S_i^{+}=\sqrt{2s-n_i}
a_i,\\ ~~~S_i^z=s-n_i; ~~~~~n_i=a_i^{\dagger }a_i \label{HP}
\end{array}
\end{equation}
The HP transformed FM hamiltonian read:
\begin{equation}\begin{array}{rl}
H=& \displaystyle
\sum_i\epsilon a_i^{\dagger }a_i-\sum_{\langle ij\rangle }J_{ij}[\frac
12(a_i^{\dagger }A_{ij}a_j+{\rm h.c.})\\ 
& \displaystyle +a_i^{\dagger }a_ia_j^{\dagger }a_j]-\frac 14\epsilon N 
\end{array} \label{FMH}
\end{equation}
and the AFM one read: 
\begin{equation}\begin{array}{rl}
H=& \displaystyle
\sum_i\epsilon a_i^{\dagger }a_i+\sum_{\langle ij\rangle }J_{ij}[\frac
12(a_i^{\dagger }a_j^{\dagger }A_{ij}+{\rm h.c.})\\ 
& \displaystyle -a_i^{\dagger }a_ia_j^{\dagger }a_j]-\frac 14\epsilon N 
\end{array} \label{AFMH}
\end{equation}
Here $N$ was the total number of the sites, $\epsilon =JZ/2$, and
$A_{ij}=\sqrt{2s-n_i}\sqrt{2s-n_j}$.
To completely remove the effect of the extra unphysical states defined by
$|m\rangle_i=a^{\dagger m}_i|0\rangle$ ($m>2s$), 
we introduced the following infinite
coupling term into the hamiltonians.
\begin{equation}
H_U=\sum_i\frac{U}{(2s+1)!}a_i^{\dagger (2s+1)}a_i^{(2s+1)}, 
~ U\rightarrow\infty    \label{HU}
\end{equation}
so that our total hamiltonians were defined by $H'=H+H_U$. 
The extra unphysical states were raised infinitely high in energy 
so that they actually decoupled from the $2s+1$ physical states.
The hamiltonians included the square roots of the operators: $\sqrt{2s-n_i}$
which was able to be expanded in terms of $a_i^{\dagger m}a_i^{m}$. The $U$ term
makes the expansion automatically truncated into a  
finite series accurately.
\begin{equation}\begin{array}{l}
\displaystyle
\sqrt{2s-n_i} = \sum_{m=0}^{2s}(-1)^mC_m\frac{1}{m!}a_i^{\dagger m}a_i^{m}\\
\displaystyle
C_m = \sum^m_i(-1)^i\frac{m!}{i!(m-i)!}\sqrt{2s-i}
\end{array} \label{sqrt}
\end{equation}
Therefore the hamiltonians only consisted of finite number of operator 
product terms.\\

\noindent {\large\it Hubbard approximation: }
Since we had the infinite coupling terms in our hamiltonians, 
an unperturbation method was needed to treat them properly. 
Our approximation was that the on-site $U$ energy hierachy was conserved
during decoupling process. This meant that we made the decoupling
$n_ia_j=\langle n_i\rangle a_j$ only if $i\not= j$ and instead we made
further equations of motion for $n_ia_i$. For half spin the resultant
3D FM magnetization and AFM sublattice magnetization are demonstrated in
Figure 1 and Figure 2, respectively. Our results are much better than
the existing spin wave theories, especially near the transition 
temperatures. At transition
temperatures our magnetization and sublattice spin tend to zero with an exponent of 1/2
but those of nonlinear spin wave theories (NLSW) are uncontinuous and 
unreasonable\cite{mattis}.
We can obtain lower ground state energy than any existing 
theories in the AFM case.
\begin{figure}[htb]
\vspace{-0.7cm}
\psfig{file=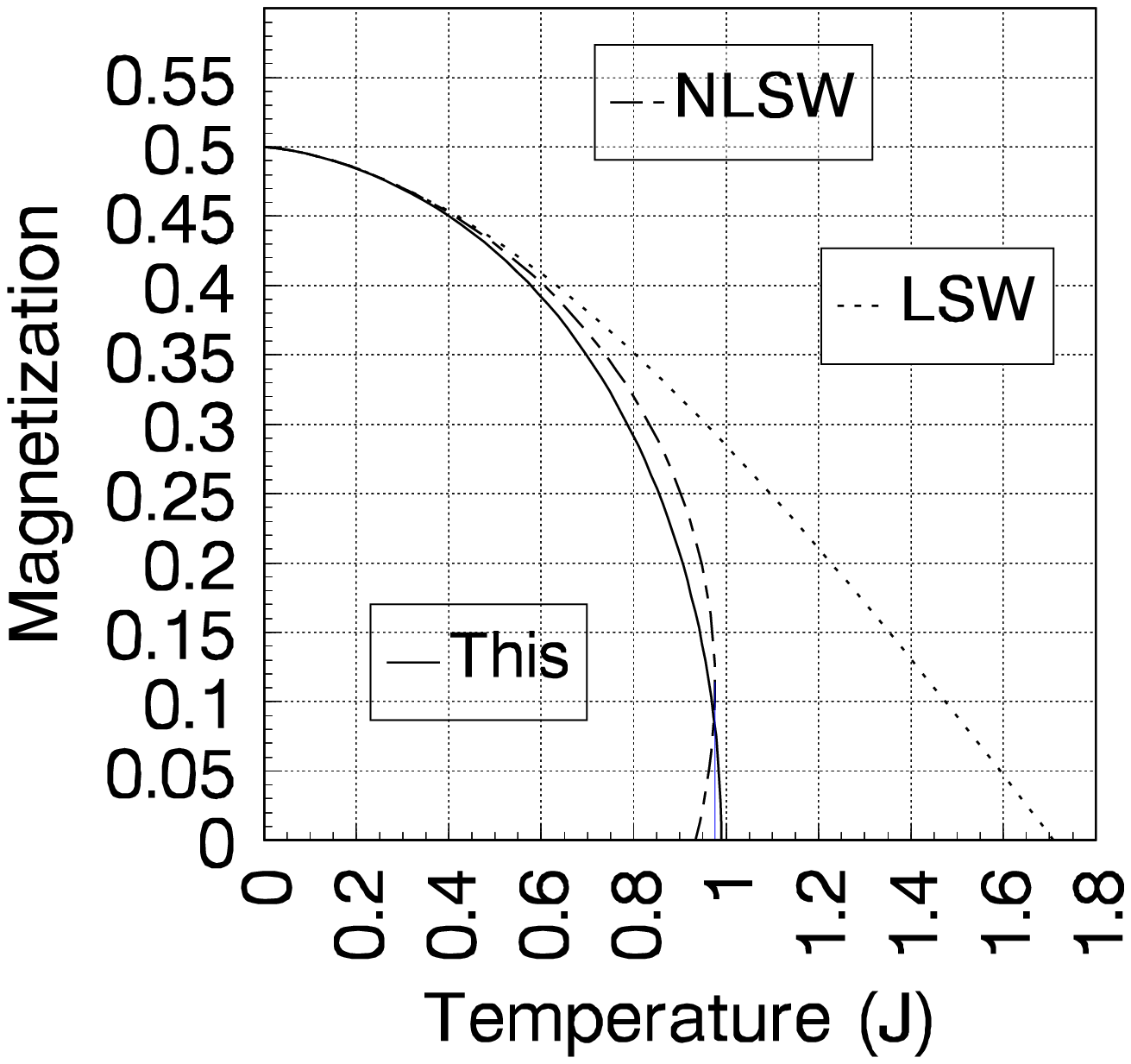,height=5.cm,width=6.5cm}
\vspace{-1cm}
\caption{3D FM magnetization compared to linear (LSW) and nonlinear 
(NLSW) spin wave theories.}
\end{figure}
\begin{figure}[htb]
\vspace{-1.4cm}
\psfig{file=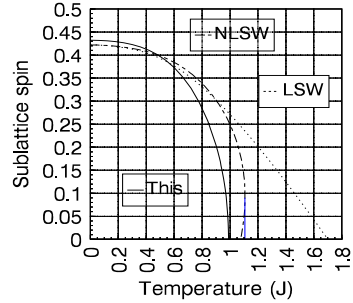,height=5.cm,width=6.5cm}
\vspace{-1cm}
\caption{3D AFM sublattice spin compared to spin wave theories. }
\end{figure}


\end{document}